\documentclass[amsmath,amssymb,twocolumn]{revtex4}
\usepackage[parfill]{parskip}    
\usepackage{graphicx}
\usepackage{amssymb}
\usepackage{epstopdf}
\usepackage{color}
\usepackage{graphicx}
\usepackage{pdfpages}

\begin{document}

\title{Minimal Geometric Deformation: the inverse problem}
\author{Ernesto Contreras {\footnote{On leave 
from Universidad Central de Venezuela}
\footnote{ejcontre@espol.edu.ec}} }
\address{
Escuela Superior Polit\'ecnica del Litoral, ESPOL, Facultad de Ciencias Naturales y 
Matem\'aticas, Apartado Postal 09-01-5863, Campus Gustavo Galindo Km 30.5 V\'ia Perimetral, Guayaquil, Ecuador.\\}
\begin{abstract}
In this paper we show that any static and spherically symmetric anisotropic solution of the Einstein
field equations can be thought as a system sourced by certain deformed isotropic system 
in the context of Minimal Geometric Deformation-decoupling approach. To be more precise, we
developed a mechanism to obtain an isotropic solution from any anisotropic solution of the Einstein field
equations. As an example, we implement the method to obtain the sources of a simple 
static anisotropic and spherically symmetric traversable wormhole.
\end{abstract}

\maketitle

\section{Introduction}\label{intro}
The Minimal Geometric Deformation (MGD) decoupling method, which
was originally developed in the context of the brane-world \cite{randall1999a,randall1999b} (see also \cite{antoniadis1990,antoniadis1998}), has been successfully
used to generate brane-world configurations from general relativistic perfect fluid solutions. Now we know that the utility of this method goes beyond this context
\cite{ovalle2008,ovalle2009,ovalle2010,casadio2012,ovalle2013,ovalle2013a,
casadio2014,casadio2015,ovalle2015,casadio2015b,
ovalle2016, cavalcanti2016,casadio2016a,ovalle2017,
rocha2017a,rocha2017b,casadio2017a,ovalle2018,estrada2018,ovalle2018a,lasheras2018,gabbanelli2018,
sharif2018,fernandez2018,fernandez2018b,contreras2018}, and a new interpretation of it \cite{ovalle2017} in terms of the so-called gravitational decoupling, is being actively used 
\cite{ovalle2017,ovalle2018,estrada2018,ovalle2018a,lasheras2018,gabbanelli2018,
sharif2018,contreras2018}. This is precisely the scenario to study in this paper.
In this sense, the MGD-decoupling represents the first systematic way to decoupling Einstein field equations, opening thus a new window in the analysis and study of these equations. By using this approach it is interesting to note how
local anisotropy can be induced in well known spherically
symmetric isotropic solutions of self–-gravitating objects. 
Even more, the isotropic solution
can be thought as a {\it generator} which, after gravitational interaction with
certain {\it decoupler} matter content, 
lead through the MGD-decoupling to an anisotropic solution of the Einstein field equations.

Based on the above statement we may wonder if it can be possible to solve the inverse problem, namely, given any
anisotropic solution of the Einstein field equations
we can study if it is possible to obtain both its {\it isotropic generator}
and its {\it decoupler} matter content.
This inverse problem program 
could be implemented, for example, in situations where an anisotropic solution is known but the mechanisms behind such a geometry are unclear. 
Such is the case of traversable static and spherically symmetric wormholes
\cite{Flamm1916,Einstein1935,Wheeler,
ellis1973,bronnikov1973,ellis1979,Morris1987,Morris1988,Visserbook,Lobo2017}
which in many cases are sourced by anisotropic fluids (for iso-\break tropic wormhole solutions see
\cite{cataldo2016}). It is well known that\break
 traversable wormhole
solutions suffer from some issues related to their instability and  violation of the
energy conditions which has been explained in terms of the so--called exotic 
matter \cite{Morris1988,Visserbook,Lobo2017}. For these reasons, it could be
interesting to explore the existence of isotropic 
solutions obeying suitable energy conditions which, after the MGD-decoupling, 
lead to wormholes geometries. Even more, it could be interesting to study if the MGD-decoupling could be considered as a kind of mechanism 
responsible for the breaking of the isotropy in the matter sector and for
the appearance of the exotic source.

In this work we develop 
a method to obtain the isotropic generator of any anisotropic solution.
More precisely, given any anisotropic solution 
we provide their sources through the MGD-decoupling (the generator 
perfect fluid and the decoupler matter content) and, as an example, we implement it 
to obtain the sources of a simple wormhole model.

This work is organized as follows. In the next section we briefly review the MGD-decoupling method.
In section \ref{isotropation} we develop the method to obtain the generator of any anisotropic solution 
of the Einstein field equations and then, we implement the method in a wormhole configuration 
in section \ref{wormhole}. The last section is devoted to final comments and conclusion.

\section{Einstein Equations and MGD--decoupling}\label{mgd}
Let us consider the Einstein field equations
\begin{eqnarray}
R_{\mu\nu}-\frac{1}{2}R g_{\mu\nu}=-\kappa^{2}T_{\mu\nu}^{tot},
\end{eqnarray}
and assume that the total energy-momentum tensor is given by
\begin{eqnarray}\label{total}
T_{\mu\nu}^{(tot)}=T_{\mu\nu}^{(m)}+\alpha\theta_{\mu\nu},
\end{eqnarray}
where $T^{\mu(m)}_{\nu}=diag(-\rho,p,p,p)$ is the matter energy momentum for a perfect fluid and $\theta_{\mu\nu}$ is an additional source coupled with the perfect fluid by the constant $\alpha$.
Since the Einstein tensor is divergence free, the total energy momentum tensor $T_{\mu\nu}^{(tot)}$
satisfies

\begin{eqnarray}\label{cons}
\nabla_{\mu}T^{(tot)\mu\nu}=0.
\end{eqnarray}

Besides, we demand that the sources do not exchange energy--momentum between them but interact only gravitationally, which implies
\begin{eqnarray}
\nabla_{\mu}T^{\mu(m)}_{\nu}=\nabla_{\mu}\theta^{\mu}_{\nu}=0.
\end{eqnarray}
In what follows, we shall consider spherically symmetric space--times with line element
parametrized as
\begin{eqnarray}\label{le}
ds^{2}=e^{\nu}dt^{2}-e^{\lambda}dr^{2}-r^{2}d\Omega^{2},
\end{eqnarray}
where $\nu$ and $\lambda$ are functions of the radial coordinate $r$ only. 
Considering Eq. (\ref{le}) as a solution of the Einstein field equations, we obtain
\begin{eqnarray}
\kappa^{2} \tilde{\rho}&=&\frac{1}{r^{2}}+e^{-\lambda}\left(\frac{\lambda'}{r}-\frac{1}{r^{2}}\right)\label{eins1}\\
\kappa^{2} \tilde{p}_{r}&=&-\frac{1}{r^{2}}+e^{-\lambda}\left(\frac{\nu'}{r}+\frac{1}{r^{2}}\right)\label{eins2}\\
\kappa^{2} \tilde{p}_{\perp}&=&\frac{e^{-\lambda}}{4}\left(\nu'^{2}-\nu '\lambda '+2\nu''
+2\frac{\nu'-\lambda'}{r}\right),\label{eins3}
\end{eqnarray}
where the primes denote derivation respect to the radial coordinate and we have defined
\begin{eqnarray}
\tilde{\rho}&=&\rho+\alpha\theta^{0}_{0}\label{rot}\\
\tilde{p}_{r}&=&p-\alpha\theta^{1}_{1}\label{prt}\\
\tilde{p}_{\perp}&=&p-\alpha\theta^{2}_{2}.\label{ppt}
\end{eqnarray} 

It is well known that solutions of equations 
(\ref{eins1}), (\ref{eins2}) and (\ref{eins3}) can be 
obtained by the MGD--decoupling method \cite{ovalle2017,ovalle2018,estrada2018,ovalle2018a,lasheras2018,gabbanelli2018,
sharif2018}. In particular, a well known isotropic system 
can be extended to, in some cases more realistic, 
anisotropic domains. The method consists in decoupling the Einstein field
equations (\ref{eins1}), (\ref{eins2}) and (\ref{eins3}) by performing
\begin{eqnarray}\label{def}
e^{-\lambda}=\mu +\alpha f,
\end{eqnarray}
where $f$ is the geometric deformation undergone by the radial metric
component $\mu$ ``controlled'' by the free parameter $\alpha$ which
is a Lorentz and general coordinate invariant \cite{ovalle2008,ovalle2009,ovalle2010,casadio2012,ovalle2013,ovalle2013a,casadio2014,
casadio2015,ovalle2015,casadio2015b,ovalle2016,cavalcanti2016,casadio2016a,ovalle2017,rocha2017a}. Doing so, we obtain
two sets of differential equations: one describing an isotropic system sourced by
the conserved energy--momentum tensor of a perfect fluid $T^{\mu(m)}_{\nu}$ an the other
set corresponding to quasi--Einstein field equations with a matter sector given by $\theta_{\mu\nu}$. More precisely, 
we obtain
\begin{eqnarray}
\kappa ^2 \rho &=&\frac{1-r \mu'-\mu}{r^{2}}\label{iso1}\\
\kappa ^2 p&=&\frac{r \mu  \nu '+\mu -1}{r^{2}}\label{iso2}\\
\kappa ^2 p&=&\frac{\mu ' \left(r \nu '+2\right)+\mu  \left(2 r \nu ''
+r \nu '^2+2 \nu '\right)}{4 r},\label{iso3}
\end{eqnarray}
for the perfect fluid and
\begin{eqnarray}
\kappa ^2  \theta^{0}_{0}&=&-\frac{r f'+f}{r^{2}}\label{aniso1}\\
\kappa ^2 \theta^{1}_{1}&=&-\frac{r f \nu '+f}{r^{2}}\label{aniso2}\\
\kappa ^2\theta^{2}_{2}&=&-\frac{f' \left(r \nu '+2\right)+f \left(2 r \nu ''+r \nu '^2+2 \nu '\right)}{4 r},\label{aniso3}
\end{eqnarray}
for the anisotropic system. It is worth noticing that the conservation equation $\nabla_{\mu}\theta^{\mu}_{\nu}=0$ leads to
\begin{eqnarray}
(\theta^{1}_{1})'-\frac{\nu'}{2}(\theta^{0}_{0}-\theta^{1}_{1})-\frac{2}{r}(\theta^{2}_{2}-\theta^{1}_{1})=0,
\end{eqnarray}
which is a linear combination of Eqs. (\ref{aniso1}), (\ref{aniso2}) and (\ref{aniso3}). In this
sense, there is no exchange of energy--momentum tensor between the perfect fluid and the
anisotropic  source as required. \\

As commented before, the MGD has been implemented to extend
isotropic solutions by performing the following protocol: given the metric 
functions $\{\nu,\mu\}$ 
and the matter content $\{\rho,p\}$ that solve Eqs. (\ref{iso1}), (\ref{iso2}) and (\ref{iso3}),
the deformation function $f$ is obtained from Eqs. (\ref{aniso1}), (\ref{aniso2}) and (\ref{aniso3})
after suitable conditions on the anisotropic source $\theta_{\mu\nu}$ 
(a linear 
or barotropic equation of state, tracelessness of the corresponding sector, etc. \cite{ovalle2018a}) are provided. 
With this information
at hand, the ``coupled'' original set of equations (\ref{eins1}), (\ref{eins2}) and
(\ref{eins3}) are finally solved. In this work we show that the scope of the MGD goes beyond the
anisotropization of isotropic solution given that it allows to solve the inverse problem. To be more precise, we use the MGD-decoupling to obtain the isotropic generator and the decoupler matter content of any anisotropic
 solution of the original equations (\ref{eins1}), (\ref{eins2}) and (\ref{eins3}) as it will be explained in the next section.

\section{MGD-decoupling: the inverse problem}\label{isotropation}
In this section we study how to obtain the isotropic generator and the
decoupler matter content of any aniso-\break tropic
system by the MGD-decoupling. The crucial point is to realize that given any anisotropic solution
with metric functions $\{\nu,\lambda\}$, matter content $\{\tilde{\rho},\tilde{p}_{r},\tilde{p}_{\perp}\}$ and definitions
in Eqs. (\ref{rot}), (\ref{prt}) and (\ref{ppt}), the following constraint must be
satisfied
\begin{eqnarray}\label{constraint}
\tilde{p}_{\perp}-\tilde{p}_{r}=-\alpha(\theta^{2}_{2}-\theta^{1}_{1}).
\end{eqnarray}
In this sense,  we can combine Eqs. (\ref{aniso2}) 
and (\ref{aniso3}) in order to obtain the geometric deformation function $f$. With this information at hand
we are able to use the geometry deformation $e^{-\lambda}=\mu+\alpha f$ to obtain
the remaining metric function of the perfect fluid $\mu$ and its matter content.
More precisely, the combination of Eqs. (\ref{aniso2}) 
and (\ref{aniso3}) with the constraint (\ref{constraint}) leads to a differential equation 
for the decoupling function $f$ given by
\begin{eqnarray}\label{f}
f'-\mathcal{F}_{1}f=\mathcal{F}_{2},
\end{eqnarray}
where we have introduced the auxiliary functions $\mathcal{F}_{1}$ and $\mathcal{F}_{2}$  as
\begin{eqnarray}
\mathcal{F}_{1}&=&\frac{4-r \left(2 r \nu ''+\nu ' \left(r \nu '-2\right)\right)}{r \left(r 
\nu '+2\right)}\\
\mathcal{F}_{2}&=&\frac{e^{-\lambda } \left(r 
\left(-\lambda ' \left(r \nu '+2\right)+2 r \nu ''+\nu ' \left(r \nu '-2\right)\right)\right.}{\alpha  r \left(r \nu '+2\right)}\nonumber\\
&& +\frac{\left.4 e^{-\lambda }\left(e^{\lambda}-1\right)\right)}{\alpha  r \left(r \nu '+2\right)}.
\end{eqnarray}
From Eq. (\ref{f}), it is straightforward that the deformation function $f$ is given by
\begin{eqnarray}\label{fs}
f(r)=e^{\int^r \mathcal{F}_{1} \, du} \int^r \mathcal{F}_{2} e^{-\int^{w}\mathcal{F}_{1} \, du} \, dw.
\end{eqnarray}
The next step consists in to obtain the metric function $\mu$ 
by replacing Eq. (\ref{fs}) in the geometric
deformation relation (\ref{def}), from where 
\begin{eqnarray}\label{mu}
\mu=e^{-\lambda}-\alpha  e^{\int^r \mathcal{F}_{1} \, du} \int^r
\mathcal{F}_{2} e^{-\int^{w} \mathcal{F}_{2} \, du} \, dw.
\end{eqnarray}

Now, from  Eqs. (\ref{iso1}), (\ref{iso2}) and (\ref{iso3}), the matter content for the 
isotropic system reads
\begin{eqnarray}
\rho&=&\mathcal{G}_{1}+\alpha  \mathcal{G}_{2} e^{\int^r \mathcal{F}_{1} \, du} 
\int^r \mathcal{F}_{2} e^{-\int^{w} \mathcal{F}_{1} \, du} \, dw\label{rho}\\
p&=&\mathcal{G}_{3}-\alpha  \mathcal{G}_{4} e^{\int^r \mathcal{F}_{1} \, du} 
\int^r \mathcal{F}_{2}e^{-\int^{w} \mathcal{F}_{1} \, du} \, dw\label{pr}.
\end{eqnarray}
where we have introduced four additional auxiliary functions 
\begin{eqnarray}
\mathcal{G}_{1}&=&\frac{r e^{-\lambda} \left(\left(e^{\lambda}-3\right) \nu '+2 r \nu ''
+r \nu '^2\right)}{8 \pi  r^2 \left(r \nu '+2\right)}\nonumber\\
&&+\frac{6 e^{-\lambda } \left(e^{\lambda }-1\right)}{8 \pi  r^2 \left(r \nu '+2\right)}
\\
\mathcal{G}_{2}&=&\frac{6-r \left(2 r \nu ''+\nu ' \left(r \nu '-3\right)\right)}{8 \pi  r^2 \left(r \nu '+2\right)}\\
\mathcal{G}_{3}&=&\frac{e^{-\lambda } \left(-e^{\lambda }+r \nu '+1\right)}{8 \pi  r^2}\\
\mathcal{G}_{4}&=&\frac{r \nu '+1}{8 \pi  r^2}.
\end{eqnarray}

To determine the decoupler matter content we simply replace Eqs (\ref{fs}) and (\ref{mu}) in (\ref{aniso1}), (\ref{aniso2}) and (\ref{aniso3}) to obtain
\begin{eqnarray}
\theta^{0}_{0}&=&
-\frac{(r\mathcal{F}_{1}+1) e^{\int^r \mathcal{F}_{1} \, du} \int^r \mathcal{F}_{2} e^{-\int^w \mathcal{F}_{1} \, du} \, dw}{r^2}\nonumber\\
&&+\frac{\mathcal{F}_{2}}{r}\label{teta00}\\
\theta^{1}_{1}&=&
-\frac{\mathcal{H}_{1} e^{\int^r \mathcal{F}_{1} \, du} \left(\int^r \mathcal{F}_{2}
 e^{-\int^w \mathcal{F}_{1} \, du} \, dw\right)}{r^2}\label{teta11}\\
\theta^{2}_{2}&=&
-\frac{ e^{\int^r \mathcal{F}_{1} \, du} \left(\int^r \mathcal{F}_{2} 
e^{-\int^w \mathcal{F}_{1} \, du} \, dw\right)}{4 r}\label{teta22}
\nonumber\\
&&+\frac{\mathcal{F}_{2} 
\left(\mathcal{H}_{1}+1\right)}{4 r},
\end{eqnarray}
where
\begin{eqnarray}
\mathcal{H}_{1}&=&1+r \nu '\\
\mathcal{H}_{2}&=&\left(\left(r \nu '+2\right) \left(\mathcal{F}_{1}+\nu '(r)\right)+2 r \nu ''(r)\right).
\end{eqnarray}

At this point the inverse problem program is completed. More precisely,
Eqs. (\ref{mu}), (\ref{rho}) and (\ref{pr}) determine the isotropic generator $\{\mu,\rho,p\}$ 
and Eqs. (\ref{teta00}), (\ref{teta11}) and (\ref{teta22})
determine the decoupler matter content $\{\theta^{0}_{0},\theta^{1}_{1},\theta^{2}_{2}\}$ once any anisotropic solution $\{\nu,\lambda,\tilde{\rho},\tilde{p}_{r},\tilde{p}_{\perp}\}$ is provided.

In the next section, as an example of the applicability of the method, we shall implement it 
to obtain the isotropic generator and the decoupler matter content corresponding to a static spherically symmetric wormhole.

\section{Static spherically symmetric wormholes}\label{wormhole}
Let us consider the static and spherically symmetric wormhole
\cite{Morris1988,Visserbook,Lobo2017} line element given by \footnote{In what
follows we sall assume $\kappa^{2}=8\pi$}
\begin{eqnarray}\label{lewh}
ds^{2}=e^{2\Phi}-\frac{dr^{2}}{1-\frac{b}{r}}-r^{2}d\Omega^{2}.
\end{eqnarray}
The metric functions $\Phi$ and $b$ are arbitrary functions of the radial coordinate
$r$. As $\Phi$ is related to the gravitational redshift, it has been 
named the redshift
function, and $b$ is called the shape function \cite{Morris1988,Visserbook,Lobo2017}. From Eq. (\ref{lewh}) we 
identify $\nu=2\Phi$ and $e^{-\lambda}=1-\frac{b}{r}$. 
The matter content associated to this geometry is given by
\begin{eqnarray}
\tilde{\rho}&=&\frac{b'}{8 \pi  r^2}\label{rhott}\\
\tilde{p}_{r}&=&-\frac{2 r (b-r) \Phi '+b}{8 \pi  r^3}\label{prtt}\\
\tilde{p}_{\perp}&=&\frac{\left(r \Phi '+1\right) 
\left(-r b'+2 r (r-b) \Phi '+b\right)}{16 \pi  r^3}\nonumber\\
&&+\frac{2 r^2 (r-b) \Phi ''}{16 \pi  r^3}.\label{pptt}
\end{eqnarray}
It is well known that the exotic matter content, as it has been usually coined,
sustaining a traversable wormhole entails the violation of the weak energy condition. However, 
in order to minimize the use of such a source we can consider a distribution of matter which falls off rapidly with radius. As an example of the above mentioned case we
can consider a simple wormhole \cite{Morris1987} characterized by
\begin{eqnarray}
\Phi&=&0\label{fi}\\
b&=&r_{0}\label{minr},
\end{eqnarray}
where $r_{0}$ is the minimum radius of the throat of the wormhole. From Eqs. (\ref{fi})
the matter content reads
\begin{eqnarray}
\tilde{\rho}&=&0\label{w1}\\
\tilde{p}_{r}&=&-\frac{r_{0}}{8 \pi  r^3}\label{w2}\\
\tilde{p}_{\perp}&=&\frac{r_{0}}{16 \pi  r^3}.\label{w3}
\end{eqnarray}
As pointed out by Morris and Thorne \cite{Morris1987}, the
parameter $\zeta=\frac{\tilde{p}_{r}}{\tilde{\rho}}-1$
quantifies the amount of exotic material
needed to sustain the wormhole. In this particular
case, although the exotic material decays rapidly with
radius, $\zeta$ is positive and huge. We may wonder if there exists some mechanism 
involving well behaved matter content (fulfilling the weak energy condition at least) which could lead to a wormhole geometry sustained
by the matter content (\ref{w1}), (\ref{w2}) and (\ref{w3}). As we shall see later, the answer is affirmative and the mechanism involved is the MGD-decoupling of a perfect fluid which satisfies all the energy conditions.

In what follows, we shall implement the inverse problem program developed in the previous section
to obtain the isotropic generator which after MGD-decoupling leads
to the wormhole geometry described above. 
Replacing $\nu=2\Phi=0$ and $e^{-\lambda}=1-\frac{r_{0}}{r}$ in
Eq. (\ref{fs}) leads to
\begin{eqnarray}\label{fss}
f&=&c_1 r^2-\frac{r_{0}}{\alpha  r},
\end{eqnarray}
where $c_{1}$ is a constant of integration with dimensions of inverse of length squared.
Combining Eqs. (\ref{fss}) and (\ref{fi}) in (\ref{mu}) we obtain the metric function
\begin{eqnarray}\label{mus}
\mu&=& 1-\alpha  c_1 r^2.
\end{eqnarray}
Replacing Eqs. (\ref{mus}) and (\ref{fi}) in (\ref{rho}), (\ref{pr}) the matter content of the
perfect fluid reads
\begin{eqnarray}
\rho&=&\frac{3 \alpha  c_1}{8 \pi }\label{pf1}\\  
p&=&-\frac{\alpha  c_1}{8 \pi }\label{pf2}.
\end{eqnarray}
Note that the system described by $\{\mu,\rho,p\}$ corresponds to the isotropic generator of the wormhole geometry. To complete the inverse problem program we obtain the decoupler matter content
which after gravitational interaction with the perfect fluid $\{\rho,p\}$ leads to the appearance of the exotic matter that sustains the wormhole geometry. Combining Eqs. 
(\ref{fss}) and (\ref{fi}) in
(\ref{teta00}),(\ref{teta11}) and (\ref{teta22}) the decoupler matter content reads
\begin{eqnarray}
\theta^{0}_{0}&=&-\frac{3 c_{1}}{8 \pi }\label{d1}\\
\theta^{1}_{1}&=&\frac{r_{0}-\alpha  c_{1} r^3}{8 \pi  \alpha  r^3}\label{d2}\\
\theta^{2}_{2}&=&-\frac{2\alpha  r^3 c_{1}+r_{0}}{16 \pi\alpha  r^3 }.\label{d3}
\end{eqnarray}

Henceforth we shall study the energy conditions fulfilled by the perfect fluid and the decoupler
matter content by setting the parameter $\alpha$ and the constant of integration $c_{1}$. 
Let us start with the case $c_{1}<0$ and $\alpha <0$. Note that 
for this choice, the perfect fluid satisfies not only the weak 
but also the strong and the dominant energy conditions. In fact,
\begin{eqnarray}
\rho\ge 0\\
\rho+p\ge0\\
\rho\ge |p|.
\end{eqnarray}
For the decoupler matter content the analysis is more subtle because the 
weak energy condition is satisfied whenever
\begin{eqnarray}
\theta^{0}_{0}+\theta^{1}_{1}\ge 0\label{r1}\\
\theta^{0}_{0}+\theta^{2}_{2}\ge 0\label{r2}.
\end{eqnarray}
It is worth noticing that the requirement in Eq. (\ref{r2}) is automatic but Eq. (\ref{r1}) leads to
\begin{eqnarray}
r^{3}\ge\frac{r_{0}}{4c_{1}\alpha}.
\end{eqnarray}
In this case a suitable choice for $c_{1}$ and $\alpha$ would be
\begin{eqnarray}
c_{1}\alpha=\frac{1}{4 r_{0}^{2}},
\end{eqnarray}
which implies
\begin{eqnarray}\label{sc}
r\ge r_{0}.
\end{eqnarray}
Remember that  $r_{0}$ is the minimum length allowed in the wormhole geometry
so Eq. (\ref{sc}) does not contradict this fact. It is worth mentioning that with the above considerations the fulfilment of
the strong and the dominant energy condition lead to $r\ge r_{0}$. In this sense, the case $\alpha<0$
and $c_{1}<0$ corresponds to a perfect fluid and a decoupler matter satisfying all the
energy conditions which after gravitational interaction leads to the appearance of the 
exotic matter sustaining the wormhole geometry. Even more, the exotic source could be thought as an effective 
matter arising as a consequence of the gravitational interaction between suitable fluids through
the MGD decoupling. 

Another case of interest could be to consider $\alpha>0$ and $c_{1}>0$. Note that for this choice the
perfect fluid satisfy all the energy conditions listed above but the decoupler matter results in an
exotic matter content. In this sense, if we insist in well behaved sources of the wormhole 
geometry, this case results unattractive. As a final case we could consider $\alpha>0$ and $c_{1}<0$ or
$\alpha>0$ or $c_{1}<0$ but, as in the previous case, these choices are inadequate if we 
desire to avoid exotic content.

\section{Conclusions}\label{remarks}
In this work we have developed the inverse problem program in the context of the Minimal Geometric Deformation method. Specifically, here we show that given any anisotropic solution of the Einstein field equations, its matter content can be decomposed into a perfect fluid sector interacting gravitationally with some decoupler matter content. As it was remarked in the manuscript, this inverse problem program could be implemented in cases where some anisotropic solution is known but the mechanism behind such a configuration remains obscure. Such is the case of 
traversable wormhole solutions where the matter sustaining it is the so called exotic matter which unavoidably violates all the energy conditions. Whether the exotic matter can be experimentally found or not, it could be interesting to provide a mechanism to explain its 
apparition. For this reason we illustrated 
how the inverse problem method works studying a simple model of a traversable wormhole space--time. 
We obtained that the exotic matter of the wormhole could be decomposed into 
a perfect fluid (the isotropic generator) and a matter (the decoupler matter) satisfying 
all the energy conditions. In this sense the emergence of the exotic matter could be thought a consequence of the gravitational interaction of reasonable (at least classically) matter content 
in the Minimal Geometric Deformation scenario.

The implementation of the method to obtain other anisotropic solutions both in General 
Relativity and in extended theories of gravity is currently under study.

\end{document}